\documentclass[journal]{IEEEtran}
\usepackage{graphicx}
\usepackage{algorithmic}
\usepackage{amsmath}
\usepackage{subfig}
\usepackage{cite}
\usepackage{balance}
\usepackage{setspace}
\usepackage{subfloat}
\usepackage{romannum}
\usepackage{caption}
\usepackage{tikz}
\tikzset{
  font={\fontsize{7.5pt}{10}\selectfont}}
\usetikzlibrary{shapes.geometric, arrows,positioning,calc}
\usetikzlibrary{intersections}
\usepackage[T1]{fontenc}
\ifCLASSINFOpdf
  
\else
  
\fi

\begin{document}

\title{ Synthesis and Optimization of Multi-Objective Multi-Output QCA Circuit using Genetic Algorithm}

\author{Mahabub Hasan Mahalat, %~\IEEEmembership{Member,~IEEE,}
        %John~Doe,~\IEEEmembership{Fellow,~OSA,}
        Mrinal Goswami, Anindan Mondal,
        ~Bibhash~Sen,~\IEEEmembership{Member,~IEEE}% <-this % stops a space
%\thanks{M. Shell was with the Department
%of Electrical and Computer Engineering, Georgia Institute of Technology, %Atlanta,
\thanks{Manuscript received April 19, 2005; revised August 26, 2015.}}

\maketitle

\begin{abstract}
The physical limitations of CMOS technology triggered several research for finding an alternative technology. QCA is one of the emerging nanotechnologies which is gaining attention as a substitute of CMOS. The main potential of QCA is its ultra low power consumption, less area overhead, and high speed. Majority and inverter gates are the basic gates in QCA, which together works as a universal logic gate to implement any QCA circuit. This paper proposes an efficient methodology for optimal QCA circuit synthesis of arbitrary multi-output boolean functions. A multi-objective genetic algorithm based approach is used to reduce worst case delay and gate count of a QCA circuit. Different importance is given to worst case delay, no. of majority and no. of inverter gates. Several efficient techniques are used in order to achieve the optimal result and reduce the computational complexity furthermore additional methodologies are used to eliminate redundancies from the final solution. Comparison of the obtained results with the existing best techniques indicates, the proposed technique outperforms in terms of worst case delay and gate count.
\end{abstract}


\IEEEpeerreviewmaketitle

\section{Introduction}

\IEEEPARstart{T}{he} traditional Complementary Metal Oxide Semiconductor (CMOS) technology is approaching towards the edge of its saturation level in terms of feature size and power efficiency, further improvement may not be possible due to physical limitations\cite{arden}. Finding an alternative technology is a prime concern of current research. For the past few years, Quantum-Dot Cellular Automata (QCA) technology has been gaining attention as an alternative to CMOS technology due to its low power overhead and area efficiency\cite{gautier}. QCA cells are the basis of any QCA device. Information transfers from cell to cell through coulombic repulsion, no current flow occurs during information passing. So QCA circuits require very less power to operate. The fundamental elements in QCA technology are majority voter (MV) gate, inverter (INV) gate, and QCA wire. The set of majority and inverter gate works as a universal logic gate to design any QCA circuit.
\par Automatic design synthesis and optimization of the layout is a well-studied topic on general AND/OR gate based logic circuits\cite{chong,Nithyananthanl}. The minimal Sum of Product (SOP) or Product of Sum (POS) expressions is generally applied to implement the optimal layout of the logic circuits. But the QCA majority logic is different than AND/OR logic, so it is difficult to apply the minimal SOP or POS expression for implementation of the optimal QCA circuit. Previously, some research work has been carried out towards the automatic generation of optimal layout of a QCA circuit. But most of them has various limitations in different measures, like no. of input variables, no. of outputs and optimization criterion. Only a few attempts have been made which try to consider all these issues. The attempts are good in their own perspective, but still they are lacking at some points especially in optimization criterion. So, it is worth to present a technique considering all these aspects for optimal synthesis of QCA circuit. Heuristic based approach seems to be a good choice in the context of the current problem.
\par Genetic Algorithm (GA) which is based on the theory of natural evolution is well known for finding the optimal solution of NP problems. GA for its ability to find a globally optimal solution and less implementation overhead is a good candidate for heuristic search\cite{chu}.
\par This paper propose a technique for the optimal synthesis of QCA circuit, especially multi-output QCA circuit based on GA. All the major optimization parameters like gate count, maximum clock delay are taken into consideration during the synthesis process. The proposed technique is applicable to single as well multi-output functions with arbitrary no. of input variables. The major contributions of this paper are summarized below
\begin{itemize}
\item An efficient fitness calculation approach is adopted for measuring the effectiveness of the candidate solutions. Importance is assigned to the different optimization criteria (gate and level) according to their cost in the QCA circuit. A relative fitness function 
is used to obtain a globally optimal solution for multi-output functions.
\item An elitism based multi-objective GA is used to find the globally optimal solution in terms of gate count and maximum clock delay. Some improved techniques are used in different steps (i.e. Creation of the initial population, crossover, mutation) of GA for better performance. Additional methodologies are applied to reduce any redundancy in the obtained solution.
\item Finally, simulation is performed on some standard functions. Comparison of the results with the available existing techniques indicates a significant improvement in terms of gate count and maximum clock delay.
\end{itemize}
        
\par The leftover paper is structured as follows. Section \Romannum{2} describes the related background materials. Section \Romannum{3} presents a brief literature review. In section \Romannum{4} the proposed technique is described.   Section \Romannum{5} contains the simulation results and comparisons. Section \Romannum{6} concludes the paper.


\section{Background Materials}
\subsection{Quantum Dot Cellular Automata}

\subsubsection{QCA Basics}
QCA is one of the emerging nanoelectronic technologies, which works based on the coulombic repulsion between electrons\cite{hennessy}. Each QCA cell contains four quantum dots and two free electrons trapped inside it. The four dots are positioned in four corners of a cell. 
In the most sTable state due to coulombic repulsion the free electrons may move into two corner quantum dots along either of the two diagonals. In this way, in sTable condition, a QCA cell may stay as either one of the two possible states. The two states are represented by $P='+1'$ and $P='-1'$ respectively, and considered as logic 1 and logic 0 respectively\cite{orlov}.

\begin{figure}
\centering
\subfloat[QCA Cell]{
\includegraphics[scale=.5]{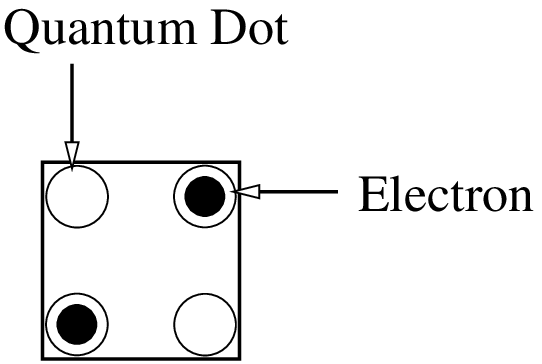}
\label{QCACell}
}
\vfill
\subfloat[QCA Logic Values]{
\includegraphics[scale=.5]{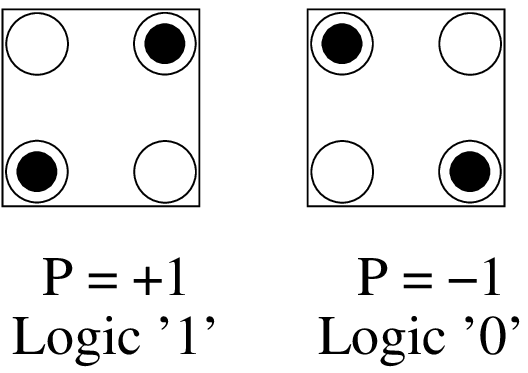}
 \label{QCAlOGIC}
     }
\caption{QCA Basics}
\label{QCA}
\end{figure}

A QCA cell and its states are shown in Fig.\ref{QCACell} and Fig.\ref{QCAlOGIC} respectively. 

The state of a QCA cell can be changed by simply applying negative or positive voltage on the cell. The signal propagates into the neighboring cells by reordering electron's position due to coulumbic repulsion.
\subsubsection{QCA Fundamental Devices}
\begin{figure}
\centering
\subfloat[Majority Voter Gate]{
\includegraphics[scale=.5]{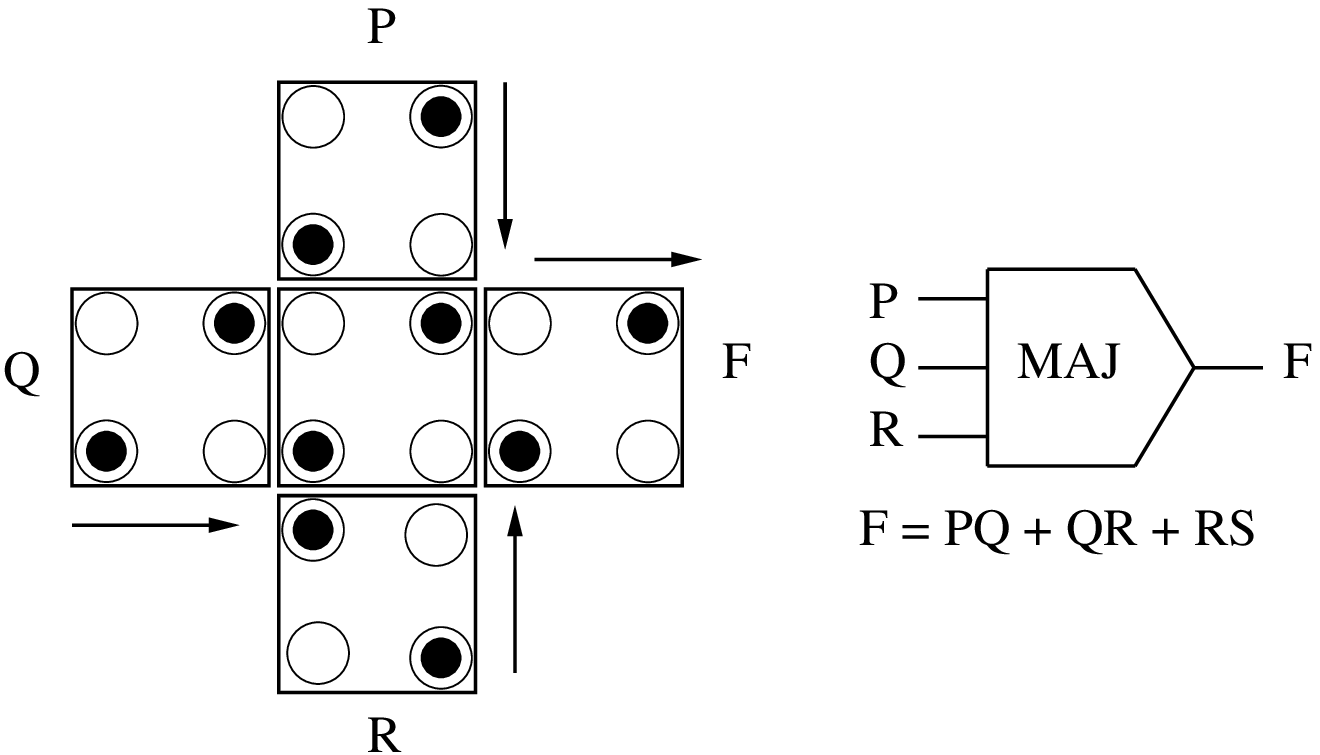}
\label{Majority}
}
\vfill
\subfloat[Different representation of Inverter Gate]{
\includegraphics[scale=.29]{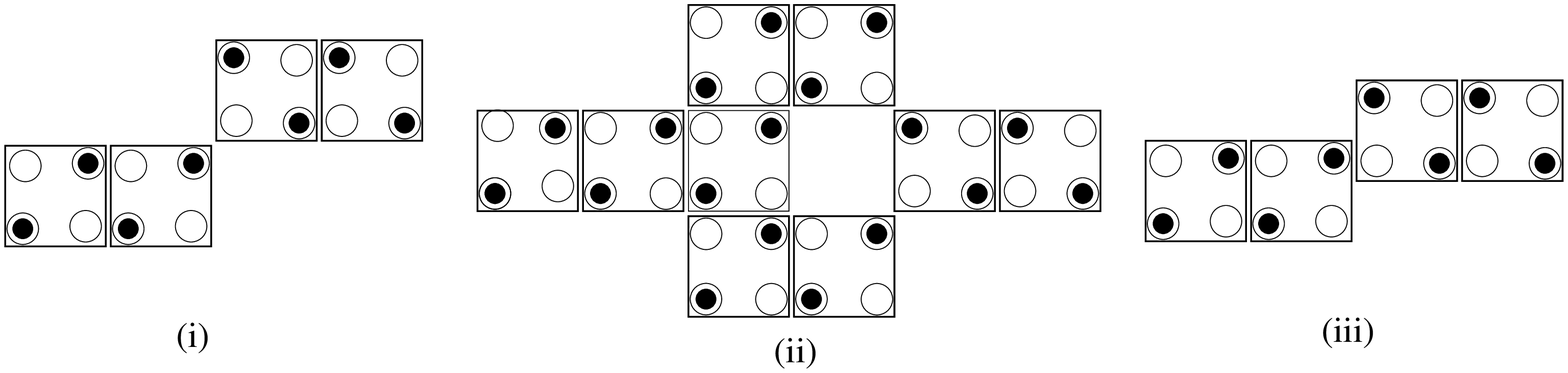}
\label{Inverter}
}
\vfill
\subfloat[QCA Wire]{
\includegraphics[scale=.3]{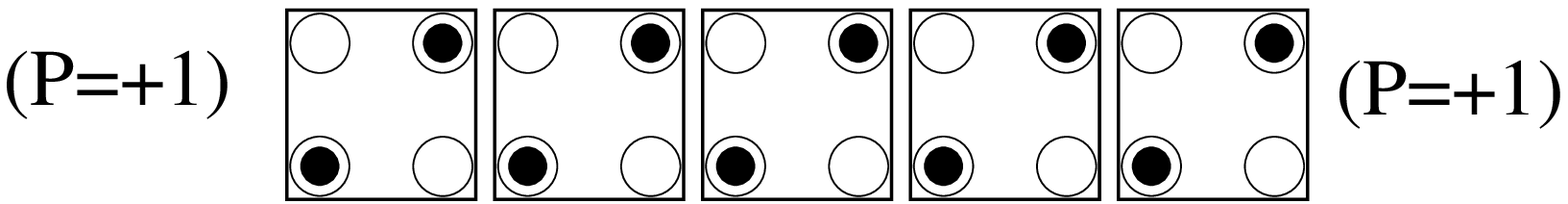}
\label{WIRE}
}
\label{QCA fundamental devices}
\caption{QCA fundamental devices}
\end{figure}

Basic QCA gates are majority voter and inverter gate. Equation of a 3 input majority gate is 
 \begin{equation}
F=PQ+QR+RP
\end{equation} 
Assuming P,Q,R as inputs and F as output. A majority voter gate is shown in Fig. \ref{Majority}. 
The output of the majority gate depends upon the majority of the inputs, it yields logic 1 as output if at least 2 of the inputs are 1 and yields logic 0 if at least two of the inputs are 0\cite{amlani}. OR gate and AND gate can be implemented using majority voter simply by fixing one of the inputs to logic 1 and logic 0 respectively. Inverter gate takes only one input and produce its complement as output. Different variants of the QCA inverter gate are shown in Fig. \ref{Inverter} ((i)-(iii)).

QCA wire is used to propagate a signal from one point of the circuit to another point. QCA wire is formed by cascading QCA cells. A QCA wire is shown in Fig. \ref{WIRE}. The signal applied to one end of a QCA wires propagates to another end through re-positioning the electrons inside the QCA cells. Any complex QCA circuit can be formed by using majority and inverter gates connecting with QCA wires.
\subsubsection{QCA Clocking}
\begin{figure}
\centering
\includegraphics[scale=.5]{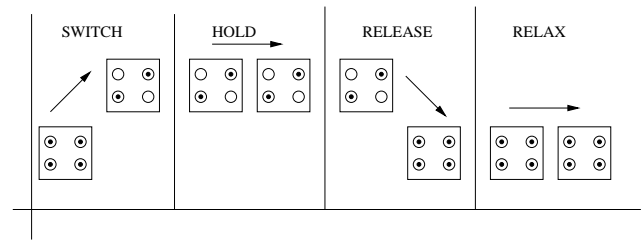}
\caption{QCA clocking}
\label{QCA clocking}
\end{figure}

Clocking is used to synchronize and control the signal flow in different parts of a QCA circuit. To get the intended result, direction and timing of signal can be controlled through clocking. A QCA clock has 4 phase switch, hold, release and relax. In switch phase, the inter-dot barrier gradually starts increasing. In hold phase QCA cell holds the same state as the applying signal. In release phase, the inter-dot barrier starts decreasing and in relax phase the cell remains in the unpolarized state. A QCA circuit is divided into several clock zones as required. Cells in the same clock zone are driven by the same clock. Increasing the number of clocks increases the worst case delay of the circuit. Phases of a QCA clock are shown in Fig.\ref{QCA clocking}

\subsection{Genetic Algorithm}
GA is based on Darwin's theory of natural evolution\cite{whitley}. This algorithm proved to be very useful for finding a feasible solution of NP problems within a limited number of computations. At first, a set of random individual solutions (chromosomes) is created, called initial population. For measuring the feasibility of each individual a fitness is assigned to each individual. Next generation population is created through selection, crossover, and mutation. The selection function selects some parent chromosomes for crossover based on a selection criterion. Crossover function is used to generate new offspring chromosomes from parent chromosomes. Mutation is introduced to maintain variations in the population pool. Mutation is performed to a chromosome by a specified probability to introduce new characteristics. The algorithm iterates several times until the goal achieved or the maximum number of iterations reached.

\section{Previous Works}
Several research papers have been proposed to synthesis and optimize conventional AND/OR-based logic circuits\cite{frenz,ali,carlos} but comparatively less number of works done to optimize QCA circuits. zhang et al.\cite{zhang} was the first to propose a method for automatic synthesis of optimal QCA circuit. The proposed method was based on Boolean Algebra and applied to generate optimal QCA circuit of 3 variable boolean functions by reducing no. of majority gates. The authors introduced 13 standard 3 variable functions, which were capable of  synthesis all the 3 variable functions. Walus et al.\cite{walus} improved the technique of \cite{zhang}. Implementation of the 13 standard functions using the improved method resulted QCA circuit with comparatively less number of majority gates. Kong et al.\cite{kong} deployed an algebraic method to systematically synthesis optimal QCA circuit for any 3 variable functions. They performed a cost analysis of a majority circuit and set priority to majority, inverter and level corresponding to their cost factor.
\par Optimization of QCA circuit using GA was first proposed by Bonyadi et al\cite{bonyadi}. A chromosome was represented using a tree structure, where internal nodes represent majority or inverter gates and leaf nodes represent constant or variable. Node count was used as a measure to reduce no. of gates in a chromosome. This method can be used for a function with an arbitrary number of inputs. Implementation of some standard functions using this technique resulted major improvements in terms of gate count. Houshmand et al.\cite{houshmand} extended the work of \cite{bonyadi} and proposed a method to reduce gate count in multi-output functions. They showed that for multi-output functions considering individual output may not lead to an optimal result, so in their technique, the circuit for n-th output function is optimized considering the previous n-1 outputs. 
\par Rezaee et al.\cite{rezaee} extended the method proposed in \cite{houshmand} to reduce both gate counts and worst case delay of multi-output functions. Optimization of worst case delay was performed by optimizing the height of the chromosome tree. To calculate the fitness value the no. of gates and the height of the chromosome tree were treated with equal importance. They did not consider any methodology to eliminate redundancies present in a  chromosome. Another major attempt to optimize multi-output QCA circuit was made by Tehrani et al\cite{tehrani}. They considered reducing gate count and worst case delay together as their objective. Their main focus was to reduce the no. of levels in the chromosome tree and totally ignored the no. of inverter gates, providing the reason that some of the implementations of inverter gate contain very few cells and it is not required to change the clock phase for an inverter (i.e inverter does not affect the speed). But the most sTable representation of inverter gate (shown in Fig. 2.b.(ii)) contains a considerable amount of cells, so though it does not affect the speed of a circuit it affects the area. However, for multi-output functions they did not synthesis the n-th output function considering previous n-1 outputs, they stored the best chromosomes for each output and the set with the highest number of common gates selected as the best global solution.

\section{Proposed Method}
It is clear from the previous works discussed in earlier section that still there are some important issues that need to be explored to synthesis the optimal solution. In the proposed work an improved technique is presented to automatically synthesis the optimal QCA circuit. The final solution is presented as a combination of majority and inverter blocks. 
\par The overview of the proposed technique is presented as a flow diagram in Fig.\ref{flowdiagram}
\tikzstyle{startstop} = [rectangle, rounded corners, minimum width=2cm, minimum height=.5cm,text badly centered, draw=black,fill=gray!30]
\tikzstyle{io} = [trapezium, trapezium left angle=70, trapezium right angle=110, minimum width=3cm, minimum height=1cm, text badly centered, draw=black]
\tikzstyle{process} = [rectangle, text width=3cm,text badly centered, draw=gray ]
\tikzstyle{decision} = [diamond,text badly centered,text width=1.8cm, aspect=2, draw=black,fill=gray!7]
\tikzstyle{line} = [draw, -latex']
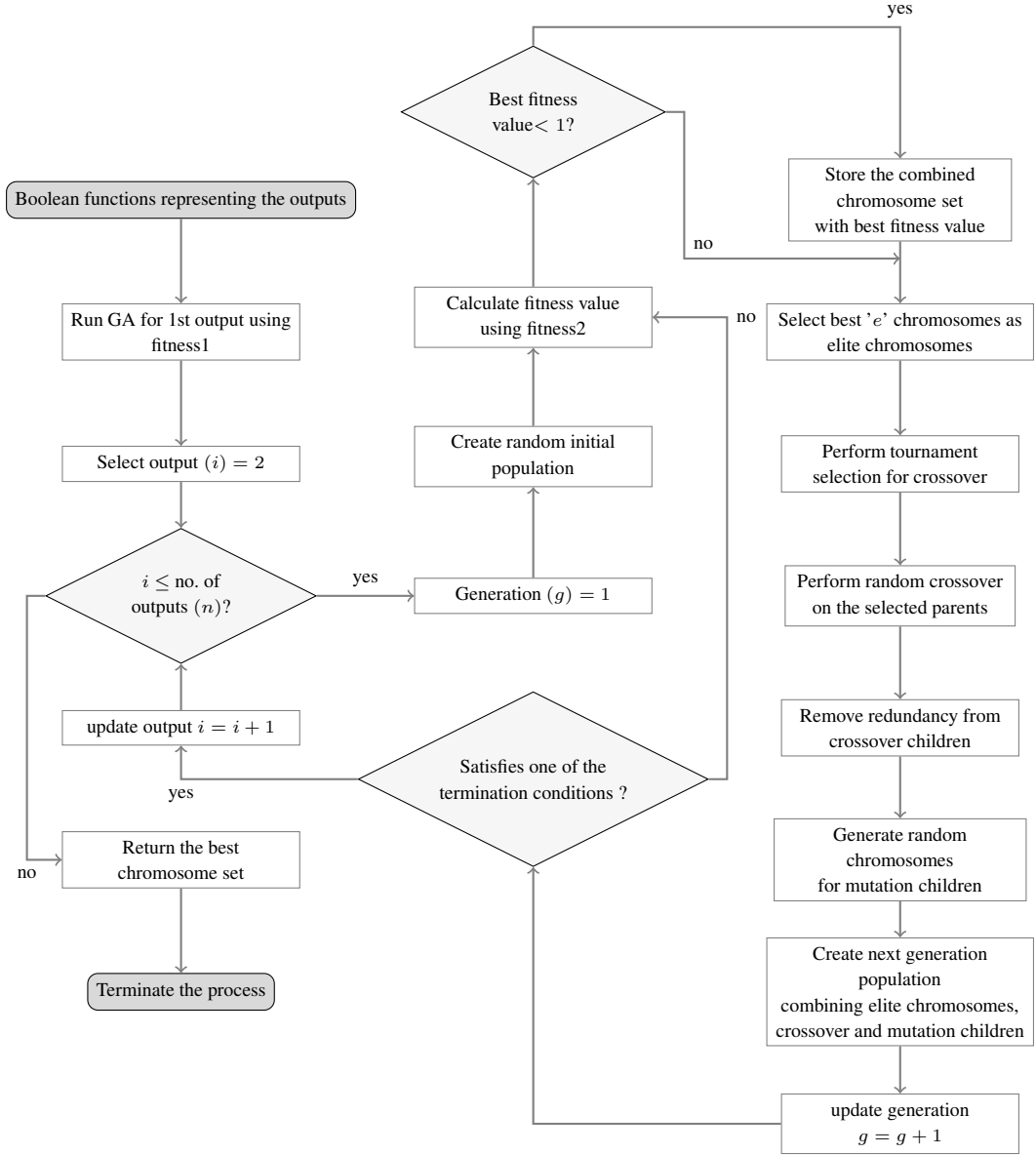
\begin{figure*}[h]

\begin{center}

\begin{tikzpicture}[node distance=1.8cm]
\node (start) [startstop] {Boolean functions representing the outputs};
%\node (in1) [io, below of=start] {Input:Output functions};
\node (pro1) [process, below of=start] {Run GA for 1st output using fitness1};
%\node (pro2) [process, below of=pro1] {Store best chromosome of each generation};
\node (pro3) [process, below of=pro1] {Select output $(i)=2$};
\node (dec1) [decision, draw=black, below of=pro3] {$i$ $\leq$ no. of outputs $(n)$?};
\node (pro2) [process, right of=dec1,xshift=3cm] {Generation $(g)=1$};
\node (pro4) [process, above of=pro2,yshift=.1cm] {Create random initial population};
\node (pro12)[process, above of=pro4,yshift=.1cm] {Calculate fitness value \\ using fitness2};
\node (dec2) [decision, above of=pro12, yshift=1cm] {Best fitness value$<1$?};
\node (pro5) [process,text width=2.8cm,right of=start, xshift=8cm] {Store the combined chromosome set \\ with best fitness value};
\node (pro6) [process, text width=3.4cm, below of=pro5] {Select  best '$e$' chromosomes as \\ elite chromosomes};
\node (pro7) [process, below of=pro6] {Perform tournament selection for crossover};
\node (pro8) [process, text width=2.9cm,below of=pro7] {Perform random crossover \\ on the selected parents};
\node (pro16) [process, below of=pro8] {Remove redundancy from crossover children};
\node (pro9) [process,text width=3.2cm, below of=pro16] {Generate random chromosomes \\ for mutation children};
\node (pro10) [process,text width=3.4cm, below of=pro9] {Create next generation population \\ combining elite chromosomes,\\ crossover and mutation children};
\node (pro13) [process, below of=pro10] {update generation \\ $g=g+1$};
\node (dec3) [decision, text width=3cm, below of=pro2, yshift=-.7cm] {Satisfies one of the\\ termination conditions ?};
\node (pro14) [process, below of=dec1] {update output $i=i+1$};
\node (pro15) [process, below of=pro14] {Return the best chromosome set};
\node (pro17) [startstop, below of=pro15] {Terminate the process};
\coordinate  (dummy1) at (9.8,-.8);
%\node (pro14) [process, below of=dec1] {};
%\node (dec4) [decision, below of=dec3, yshift=-1cm, xshift=3cm ] {n < no. of output};
%\node (pro11) [process, below of=dec4,yshift=-1cm] {mutation};
%\node (stop) [startstop, left of=pro11, xshift=-4cm] {Stop};

%\draw [thick,->,gray] (dec3) -| node[anchor=south,black] {yes} (dec4);

%\draw [thick,->,gray] (dec3) -|  (stop);
%\draw [thick,->,gray] (dec4) -- node[anchor=east,black] {yes} (pro11);
%\draw [thick,->,gray] (dec4.west) -- ++(-4.4cm,0);
\draw [thick,->,gray] (dec1) -- node[anchor=south,black] {yes} (pro2);
\draw [thick,->,gray] (pro12) --(dec2);
\draw [thick,->,gray] (dec2.north) |- ++ (.25cm,.25) -| node[anchor=south,black] {yes} (pro5);
\draw [thick,->,gray] (dec2.east) -- ++ (.25cm,0)|- node[anchor=south west,black]  {no}(dummy1);
\draw [thick,->,gray] (start) -- (pro1);
%\draw [thick,->,gray] (in1) -- (pro1);
\draw [thick,->,gray] (pro2) -- (pro4);
\draw [thick,->,gray] (pro1) -- (pro3);
%\draw [thick,->,gray] (pro2) -- (pro3);
\draw [thick,->,gray] (pro3) -- (dec1);
\draw [thick,->,gray] (pro4) -- (pro12);
\draw [thick,->,gray] (pro5) -- (pro6);
\draw [thick,->,gray] (pro6) -- (pro7);
\draw [thick,->,gray] (pro7) -- (pro8);
\draw [thick,->,gray] (pro8) -- (pro16);
\draw [thick,->,gray] (pro16) -- (pro9);
\draw [thick,->,gray] (pro9) -- (pro10);
\draw [thick,->,gray] (pro10) -- (pro13);

\draw [thick,->,gray] (pro15) -- (pro17);
\draw [thick,->,gray] (pro13.west) -| (dec3);
\draw [thick,->,gray] (dec3.east)-- ++ (.25,0) |- node[anchor=west,black] {no}(pro12);
\draw [thick,->,gray] (dec3.west) -| node[anchor=north,black] {yes}(pro14);
\draw [thick,->,gray] (pro14) --(dec1);
\draw [thick,->,gray] (dec1.west) -- ++ (-.25,0)|- node[anchor=north,black]{no}(pro15.west);
\end{tikzpicture}
\caption{Major steps of the proposed technique}
\label{flowdiagram}
\end{center}
\end{figure*}

Each output of a multi-output function is considered as an individual function, for example for a full adder two functions are required one for sum and another for carry. To apply the process first the output functions are ranked randomly as 1,2,3...N. The process begins with generating some random individual solutions each representing a valid QCA circuit. A fitness value is assigned to each individual for measuring the optimality of the solution. Some individual solutions with best fitness are selected as elite chromosomes to pass into the next generation without any alteration. Using tournament selection some parents are selected for crossover. A variant of uniform crossover is used to a pair of parents for generating child chromosome. Finally the leftover chromosomes are generated through mutation to keep the population size constant. Combination of elite chromosomes, crossover children and mutation children form the next generation population pool. This process repeats several times until the termination condition is reached. For multi output circuit the whole process is repeated for each output, distinct set of rules is used for fitness calculation of first output function and  for the functions other than first output. The process returns the best combined solution satisfying all the outputs. Details of each step of the proposed approach are discussed below

\subsection{ Creation of Initial Population}
\begin{figure}[h]
\centering
\includegraphics[scale=.5]{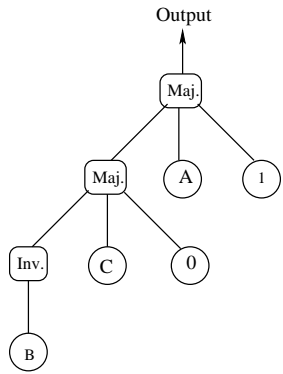}
\caption{Tree Representing Chromosome $M(M(B',C,O)A,1)$}
\label{Chromosome1}
\end{figure}

An individual chromosome (solution) is represented using the conventional tree structure\cite{bonyadi}. Internal nodes of the tree represent the majority or inverter gate and leaf nodes represent constant (i.e. 0 or 1). Pre-order traversal of the chromosome tree is used to store the chromosome. Fig. \ref{Chromosome1} represents the chromosome corresponding to the circuit
\begin{equation}
M(M(B',C,O)A,1)
\end{equation}
To generate an individual chromosome first the majority, inverter and input variables are mapped into distinct integers (i.e. $MV\rightarrow 7,Inv\rightarrow 6, A\rightarrow 5$ etc.). A random string using the set of mapped integers is generated through following the rules
\begin{enumerate}
\item To start the process an integer from the set of  mapped integers is chosen at random.
\item If the current number denotes a majority gate then 3 additional numbers are added following the current number in the sequence, if it denotes an inverter gate then one additional number is added following the current number. This rule is recursively applied to each number generated in the string. If the current number denotes an input variable or constant, then no more additional number is required.  
\item If the string length is greater then a specified length than it is discarded and a new one is generated.
\end{enumerate}
Validating the above rules bound the generated string to represent a valid QCA circuit. The string of integer is mapped into the corresponding QCA circuit.
\par For creating the set of initial population a set of chromosomes sized 10 times  of the population size is generated first. Structurally Similar chromosomes are removed from the set to maintain variations. From the chromosome set, number of chromosomes same as the population size is randomly selected to fill the set of initial population. 
\subsection{Fitness Calculation}
To calculate the fitness value the target function is represented by its canonical form. For example the 3 variable SOP expression $AB + B'C$ is represented in its canonical form as $ABC + ABC' + AB'C + A'B'C$. As mentioned earlier, two distinct approaches are used to calculate fitness value, one for the first output function and another for the output functions other than first output. 
\par Fitness value calculation for first output (fitness1) is represented in Fig. \ref{fitness1}, same algorithm can be used for single output functions.
\begin{figure}
{\fontsize{8}{8}\selectfont
\algsetup{indent=1em}
\begin{algorithmic}[1]
\STATE \textbf{Input:} Chromosome\vspace*{.08in}
\STATE \textbf{Return:} Fitness Value\vspace*{.08in}
\STATE $temp\_fit \leftarrow \frac{\text {No. of possible inputs}}{\text{No. of similar outputs}}$\vspace*{.04in}
\IF{$temp\_fit == 1$}\vspace*{.04in} 
\STATE $fitness \leftarrow 1 - (\frac{1}{\text{No. of gates}} + \frac{1}{\text{No. of levels}})$\vspace*{.04in}
\ELSE\vspace*{.04in}
\STATE $fitness \leftarrow temp\_fit$\vspace*{.04in}
\ENDIF\vspace*{.04in}
\RETURN $fitness$
\end{algorithmic}
}
\caption{Pseudo code for fitness1}
\label{fitness1}
\end{figure}
The fitness function has two goals, the first goal is to synthesize a QCA logic circuit corresponding to the output function. To achieve this goal, the fitness function assigns better fitness value to a chromosome, which is closer to the valid solution. In our convention lower fitness value implies better solution. Fitness value is calculated as the ratio of no. of possible inputs using the input variables (i.e for n variable function no. of possible inputs$=2^n$) and the no. of outputs similar to the target output function. A fitness value of 1 (i.e. $'temp\_fit'=1$) indicates the chromosome represents a valid solution.
\par The second goal of the fitness function is to synthesis the optimal circuit. Once the valid solution is generated, the fitness function tries to minimize the no. of the gates and worst case delay of the circuit. Minimization of worst case delay is realized by reducing no. of levels in the circuit. No. of level is measured in terms of  the no. of majority along the longest path from root to leaf node of the chromosome tree. Majority gate has 3 inputs, one output, so, synchronization is required for proper functioning of the gate, whereas inverter gate, having only one input and one output is not required to synchronize. Thus the inverter gate does not have much impact on delay. So inverter gates are not considered for counting the level. Increasing the level indicates increase in delay as well as area, [i.e. It increases no. of majority] which implies no. of levels is the most important factor to minimize. It was shown in \cite{kong}, for optimization of QCA circuit, level is the most important factor followed by no. of majority. No. of inverter has lowest importance. The same is followed in the proposed technique. The fitness function implicitly assigns more importance to level than no. of gates. As level is calculated in terms of no. of the majority along the longest path of a QCA circuit, so no. of level cannot be more than the no. majority in a chromosome and in most of the practical circuit no. of level is less than no. of gates in a chromosome. So the value of the  $'fitness'$ (Fig. \ref{fitness1}) is more sensitive to the no of level than no. of gates. 
\begin{figure}
{\fontsize{8}{8}\selectfont
\algsetup{indent=1em}
\begin{algorithmic}[1]
\STATE \textbf{Input:} Chromosome\vspace*{.08in}
\STATE \textbf{Return:} Fitness Value\vspace*{.08in}
\STATE $temp\_fit \leftarrow \frac{\text {No. of possible inputs}}{\text{No. of similar outputs}}$\vspace*{.04in}
\IF{$temp\_fit == 1$}\vspace*{.04in}
\IF {\textit{fitness value present in} $stored\_fitness$}\vspace*{.04in}
\STATE $fitness \leftarrow stored\_fitness(Chromosome)$\vspace*{.04in}
\ELSE \vspace*{.04in}
\STATE $final\_gates \leftarrow \infty$ \vspace*{.04in}
\STATE $final\_level \leftarrow \infty$ \vspace*{.04in}
\FOR{$i=1$ \TO \textit{no. of entry in} $stored\_chromosomes$}\vspace*{.04in}
\STATE $total\_majority\leftarrow$\textit{No. of unique majority gates in $stored\_chromosomes(i)-$ No. of common majority gates with current chromosome}\vspace*{.04in}
\STATE $total\_inverter\leftarrow$\textit{No. of unique inverters in $stored\_chromosomes(i)-$ No. of common inverters with current chromosome}\vspace*{.04in}
\STATE $total\_gates\leftarrow total\_majority+total\_inverter/3$\vspace*{.04in}
\STATE $max\_level \leftarrow$\textit{max(no of levels in current chromosome, levels in $stored\_chromosomes(i)$)}\vspace*{.04in}
\IF{$(total\_gates<final\_gates$ \AND $max\_level \leq final\_level$) \OR  $(max\_level < final\_level$)}\vspace*{.04in}
\STATE $final\_gates\leftarrow total\_gates$
\STATE$final\_level \leftarrow max\_level$
\STATE $final\_pos\leftarrow i$\vspace*{.04in}
\ENDIF\vspace*{.04in}
\ENDFOR\vspace*{.04in}
\STATE replace the common parts between the current chromosome and $stored\_chromosomes(final\_pos)$ with the common parts having minimum no. of gates.\vspace*{.04in}
\STATE $level\leftarrow$\textit{max(no of levels in current chromosome, levels in $stored\_chromosomes(final\_pos)$)}\vspace*{.04in}
\STATE$no\_of\_gates\leftarrow$ \textit{no. of unique gates in $stored\_chromosomes(final\_pos)$ + no of gates in current chromosome $-$ no of gates common with current chromosome}\vspace*{.04in}
\STATE $fitness \leftarrow 1 - (\frac{1}{level} + \frac{1}{no\_of\_gates})$\vspace*{.04in}
\STATE $stored\_fitness(Chromosome)\leftarrow fitness$\vspace*{.04in}
\ENDIF\vspace*{.04in}
\ELSE\vspace*{.04in}
\STATE $fitness \leftarrow temp\_fit$\vspace*{.04in}
\ENDIF\vspace*{.04in}
\RETURN $fitness$
\end{algorithmic}
}
\caption{Pseudo code for fitness2}
\label{fitness2}
\end{figure}
\par For the output functions other than the first output a slightly modified version of the fitness function is used. The algorithm is named as fitness2 and represented in Fig.\ref{fitness2}. To synthesis a valid chromosome corresponding to the output function, same steps as fitness1 is followed. The algorithm differs from fitness1 in next steps. A variable '$stored\_fitness$' is maintained throughout the process to store fitness of chromosomes with fitness value less than 1. In future, if the same chromosome is used again, its fitness value is just returned from $'stored\_fitness'$. This strategy reduces the computation overhead. If the current chromosome represents a valid circuit ($'temp\_fit=1'$) and it's fitness value is not present in the $'stored\_fitness'$ then the for loop is executed to calculate the fitness value. A matrix variable $'stored\_chromosomes'$ stores chromosome from previous outputs. Suppose the rank of current output function is i, then each row entry of the $'stored\_chromosomes'$ stores a chromosome set from 1 to i-1 th output, information about the no. of unique majority and inverter gates and maximum level of the chromosome set. The i th output function is optimized considering the previous i-1 outputs. For common gates between the output functions instead of repeating the common parts, both the functions can use the common parts. The common parts are checked in terms of their K-map value. To select a chromosome set of previous outputs for combining with the current chromosome, the no. of majority and inverter gets along the common parts of the previous chromosome set is subtracted from no. of majority (i.e $total\_majority$) and inverter (i.e $total\_inverter$) respectively. As circuit of an individual output function is considered as a part of the single combined circuit, so all the circuits have to operate under the same clocking scheme. Thus the worst case delay of the combined circuit depends upon the maximum worst case delay among all individual circuits. So maximum level among all the output functions is considered to measure worst case delay. 

\par
\begin{figure}[!ht]
     \subfloat[Chromosomes before combination\label{before_combine}]{%
       \includegraphics[width=0.45\textwidth]{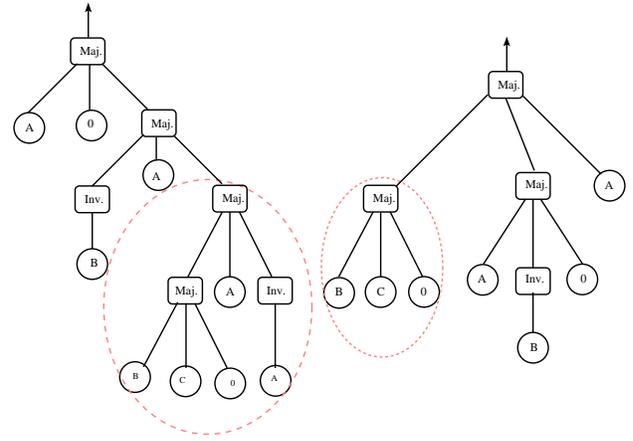}
     }
     \hfill
     \subfloat[Chromosomes after combination\label{after_combine}]{%
       \includegraphics[width=0.35\textwidth]{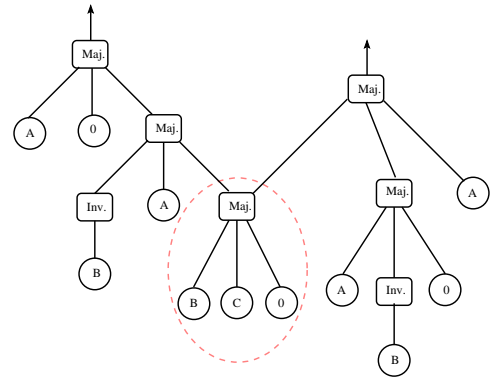}
     }
     \caption{Combination process of a two output function}
     \label{combination}
   \end{figure}

 For choosing the chromosome set of previous outputs to combine with current chromosome maximum pressure is given to level. The set that minimizes the maximum level is chosen first. No. of majority and inverter is used to break a tie among the chromosome sets having a similar maximum level. No. of the inverter is given as one-third of the priority of no. of the majority as the majority has three inputs and one output and it requires to synchronize \cite{kong}. The no. of $'total\_gates'$ is counted as the summation of no. of the majority with the one-third of the no. of the inverter. The combined chromosome set providing a minimum value of $'total\_gates'$ is selected among the chromosome set having a similar maximum level. Once a previous chromosome set is finally selected, the common parts between the current chromosome and the previous chromosome set are replaced with the common part having minimum no. of gates. Then the no. of total gates and maximum level are calculated to determine the fitness value. The combination process for a two output function is shown in Fig. \ref{combination}.
As $M(M(B,C,0),A,A')$ and $M(B,C,0)$ has same K-map value but the 2nd one having less number of gates is selected to form the combined chromosome set.

\subsection{Selection of chromosomes}
An elitism based approach is used to preserve the chromosomes with the best fitness. The chromosomes are sorted according to their fitness value and from the sorted list a small number of chromosomes (i.e. the number is predetermined) with better fitness value are forwarded into the next generation without any alteration. The tournament selection method is used for selecting parent chromosomes for crossover. No. of chromosomes similar to tournament size is selected at random. From the selected chromosomes the chromosome with best fitness value is finally picked. The tournament selection method is performed several times until the parent chromosome list for crossover is filled.  
\subsection{Crossover}
\begin{figure}[!ht]
     \subfloat[Chromosomes before crossover\label{before crossover}]{%
       \includegraphics[width=0.4\textwidth]{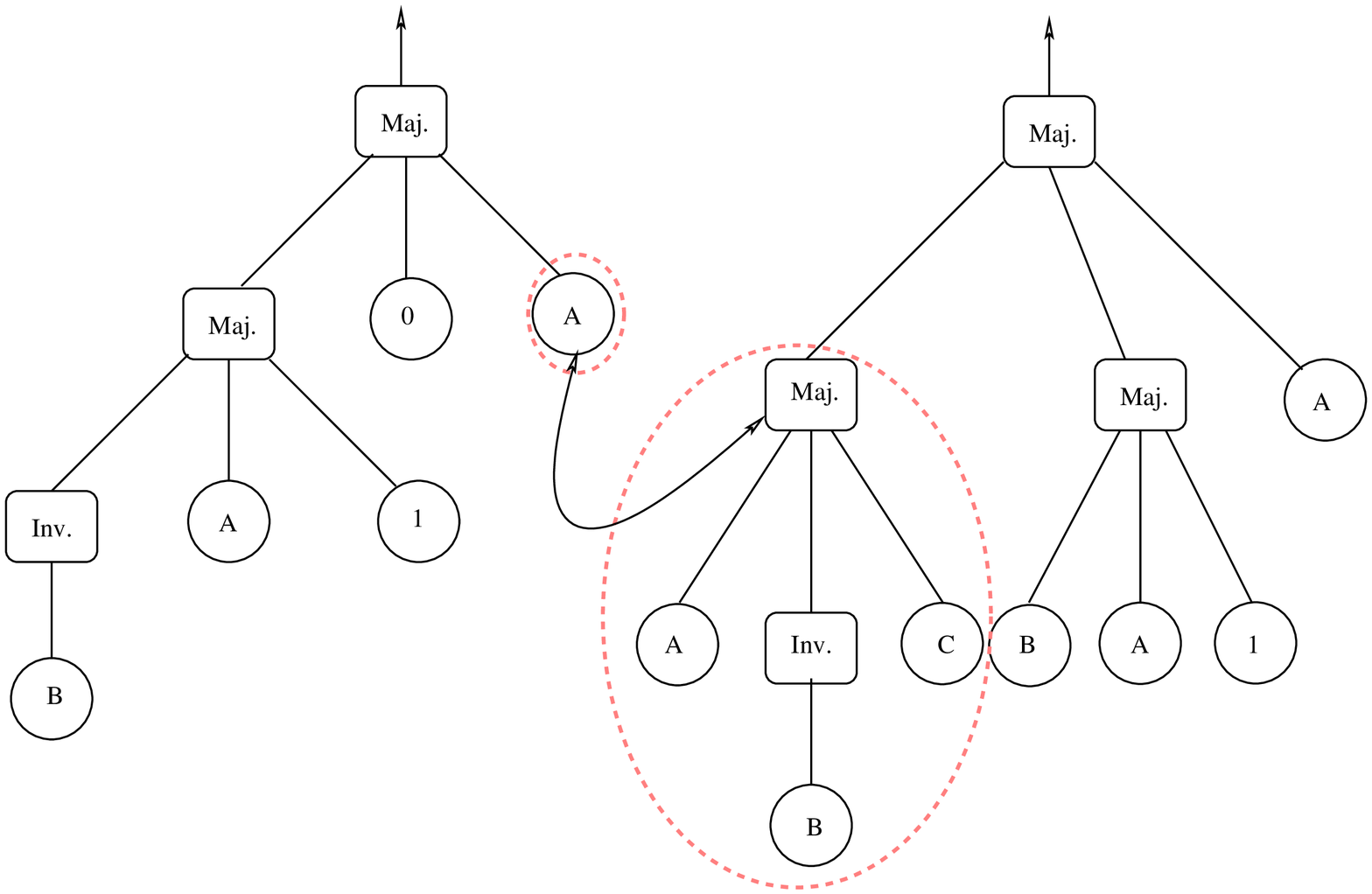}
     }
     \hfill
     \subfloat[Chromosomes after crossover\label{after crossover}]{%
       \includegraphics[width=0.4\textwidth]{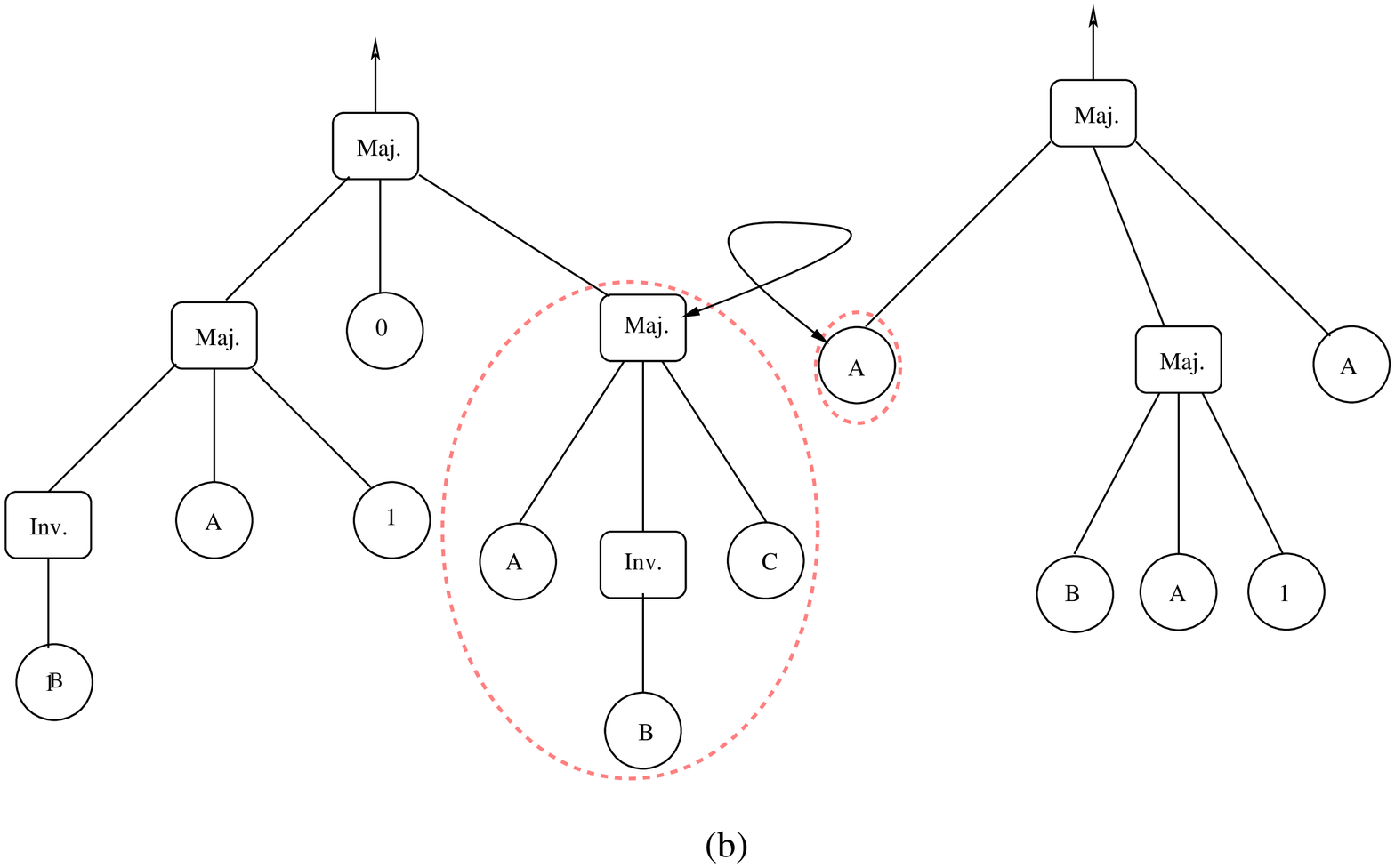}
     }
     \caption{Crossover operation between two parents}
     \label{fig:dummy}
   \end{figure}

No. of Parent chromosomes twice of the size of crossover children is selected through selection method. From the parent chromosome list, two parents are selected at random to generate a single child chromosome. To perform crossover a node from parent1 and another node from parent2 are selected randomly. The selected nodes are exchanged along with their subtree to generate two offspring. Finally, the offspring with best fitness value is taken as child chromosome. n offspring chromosomes are generated from 2n parent chromosomes. Crossover operation between two parents is shown in Fig. \ref{fig:dummy} (a-b). 
\subsection{Mutation}
As the number of inputs in majority gate and inverter gate are different, so it is very difficult to use conventional mutation functions, because replacing an inverter gate with majority or reverse may lead to an invalid QCA circuit. So a different approach is used for mutation. The mutation children are not generated from the current population, they are generated externally and added to the next generation population set. At first a large set of chromosomes is generated randomly using the same procedure used for initial population. From the generated chromosome set, number of chromosomes equal to the no. of mutation children  are selected at random and forwarded to the  next generation population pool as mutation children.
\par The complete algorithm is shown in Fig.\ref{Complete algo}. After Crossover another step is added called '$local\_improvement$' to remove any redundancy from the generated offspring. This step tries to remove redundant inverter gates and try to replace majority gate with variable or constant having similar functionality. For example, replace $M(A,B,1')$ by $M(A,B,0)$, replace $M(A,A,1)$ by $A$. 
\begin{figure}
{\fontsize{8}{8}\selectfont
\algsetup{indent=1em}
\begin{algorithmic}[1]
\STATE \textit{run GA method for output1 and stores best chromosomes of each generation in} $stored\_chromosomes$\vspace*{.04in}
\FOR{$output$ $2$ \TO $N$ }\vspace*{.04in}
\STATE \textit{generate initial population of size $pop$}\vspace*{.04in}
\FOR{\textit{generation} $1$ \TO $max\_gen$ }\vspace*{.04in}
\STATE \textit{calculate fitness of each chromosome using} $fitness2$\vspace*{.04in}
\IF{\textit{fitness of best chromosome}$<1$}\vspace*{.04in}
\STATE \textit{add the best chromosome along with corresponding previous set of chromosomes to list} $temp\_stored\_chromosome$\vspace*{.04in}
\ENDIF\vspace*{.04in}
\STATE \textit{select $e$ best chromosomes as elite chromosomes}\vspace*{.04in}
\STATE \textit{perform tournament selection to select $2c$ parent chromosomes}\vspace*{.04in}
\STATE \textit{apply crossover to generate $c$ child chromosomes from $2c$ parents}\vspace*{.04in}
\STATE \textit{apply $local\_improvement$ on each child chromosomes}\vspace*{.04in}
\STATE \textit{perform mutation to generate $pop-(c+e)$ chromosomes}\vspace*{.04in}
\IF{\textit{best fitness $<1$} \AND \textit{no changes in best fitness since $thresh\_gen$ no. of generations}}\vspace*{.04in}
\IF{$output==N$}\vspace*{.04in}
\STATE \textbf{return} \textit{best chromosome along with corresponding previous chromosomes set}\vspace*{.04in}
\STATE \textit{terminate the process}\vspace*{.04in}
\ELSE \vspace*{.04in}
\STATE $stored\_chromosomes \leftarrow temp\_stored\_chromosome$\vspace*{.04in}
\STATE \textit{delete the entries of $temp\_stored\_chromosome$}
\STATE \textit{terminate the inner for loop}\vspace*{.04in}
\ENDIF
\ENDIF
\ENDFOR
\ENDFOR
\end{algorithmic}}
\caption{The complete pseudo code for multi output function }
\label{Complete algo}
\end{figure}

The algorithm is itself self-explanatory. The outer for loop iterates for the 2nd to last output. The inner for loop generates output function for ith output relative to i-1 outputs. The combined best chromosomes of ith output with i-1 outputs are stored in $stored\_chromosomes$. The variable $stored\_chromosomes$ updated after calculation of each output. The inner for loop iterates till the maximum no of generations reached or after generating a correct chromosome no improvement in terms of best fitness is found throughout the 'thresh\_gen' no. of consecutive generations. After the last output, the outer for loop breaks and returns the combined best chromosome satisfying all the outputs.

\begin{table*}[h]
\footnotesize
%% increase Table row spacing, adjust to taste
\renewcommand{\arraystretch}{1.4}
\setlength{\tabcolsep}{.1em}
% if using array.sty, it might be a good idea to tweak the value of
% \extrarowheight as needed to properly center the text within the cells
\begin{center}
\caption{Simulation result and comparison of a 3 input/2 output function}
\label{Table1}
%% Some packages, such as MDW tools, offer better commands for making Tables
%% than the plain LaTeX2e tabular which is used here.
\begin{tabular}{|c|c|c|c|c|c|c|c|c|c|c|c|}
\hline
Approach & Minterms & Circuit & NMV  & NINV   & Levels & CMV & CINV & TMV & TINV &TG &Max  \\
         &           &        &       &       &        &      &     &      &  & &level\\\hline
\cite{tehrani}  & $F_{1}=\sum_{}{} m_{0},m_{2},m_{4},m_{7}$ & $F_{1}=$M(M(1,A,C)',M(A,B',C),M(1,B,C')) &4  & 3 & 2 & 1 & 1 & 6 & 4 & 10 & 2\\
         & $F_{2}=\sum_{}{} m_{0},m_{2},m_{3},m_{4}$ & $F_{2}=$M(0,M(0,A,B)',M(1,B,C')) &3  & 2 & 2  & & & & & &\\\hline
Proposed & $F_{1}=\sum_{}{} m_{0},m_{2},m_{4},m_{7}$ & $F_{1}=$M(M(A,C,B)',M(A,C',1),M(B,0,C)) &4  &\textbf{2}  & 2  & 1 & 0 & 6 & 4 & 10 &2\\
      Approach   & $F_{2}=\sum_{}{} m_{0},m_{2},m_{3},m_{4}$ & $F_{2}=$M(M(O,B,C)',M(B,O,A),C)'             &3  &2  &2   & & & & & &\\\hline
      
\end{tabular}
\end{center}
\end{table*} 
\begin{table*}[!h]
\footnotesize
%% increase Table row spacing, adjust to taste
\renewcommand{\arraystretch}{1.4}
\setlength{\tabcolsep}{.1em}
% if using array.sty, it might be a good idea to tweak the value of
% \extrarowheight as needed to properly center the text within the cells
\begin{center}
\caption{Simulation result and comparison of a 4 input/2 output function}
\label{Table2}
%% Some packages, such as MDW tools, offer better commands for making Tables
%% than the plain LaTeX2e tabular which is used here.
\begin{tabular}{|c|c|c|c|c|c|c|c|c|c|c|c|}
\hline
Approach & Minterms & Circuit & NMV  & NINV   & Levels & CMV & CINV & TMV & TINV &TG &Max  \\
         &           &        &       &       &        &      &     &      &  & &level\\\hline
\cite{rezaee} & $F_{1}=\sum_{}{} m_{0},m_{2},m_{6},m_{12},$ & $F_{1}=$M(M(A,1',C)',M(D',M(B,A,1)', 
 & 6 & 5 & 3 & 3 & 3 & 8 & 5 & 13 & 3 \\
       & $m_{13},m_{14}$ & M(A,B,C)),M(1',B,A)) & & & & & & & & &\\
       & $F_{2}=\sum_{}{} m_{1},m_{3},m_{4},m_{5},$ & $F_{2}=$M(B,M(A,C,1')',M & 5 & 3 & 3 & & & & & &\\
         &$m_{7},m_{12},m_{13},m_{15}$ &(M(B,A,1)',D,M(A,B,C))) & & & & & & & & &\\\hline

Proposed & $F_{1}=\sum_{}{} m_{0},m_{2},m_{6},m_{12},$ & $F_{1}=$M(M(B,1,A'),M(M(B,C,1)',C,D'), 
 & 6 & \textbf{4} & 3 & 2 & 2 & \textbf{7} & \textbf{4} & \textbf{11} & 3\\
 Approach & $m_{13},m_{14}$ & M(A,0,M(0,B,C'))) & & & & & & & & &\\
         & $F_{2}=\sum_{}{} m_{1},m_{3},m_{4},m_{5},$ & $F_{2}=$M(D,M(1,A',B),M(C',0,B)) & \textbf{3} & \textbf{2} & \textbf{2} & & & & & &\\
         &$m_{7},m_{12},m_{13},m_{15}$ & & & & & & & & & &\\\hline
      
\end{tabular}
\end{center}
\end{table*}
\begin{table*}[!h]
\footnotesize
%% increase Table row spacing, adjust to taste
\renewcommand{\arraystretch}{1.4}
\setlength{\tabcolsep}{.1em}
% if using array.sty, it might be a good idea to tweak the value of
% \extrarowheight as needed to properly center the text within the cells
\begin{center}
\caption{Simulation result and comparison of a 3 input/3 output function}
\label{Table3}
%% Some packages, such as MDW tools, offer better commands for making Tables
%% than the plain LaTeX2e tabular which is used here.
\begin{tabular}{|c|c|c|c|c|c|c|c|c|c|c|c|}
\hline
Approach & Minterms & Circuit & NMV  & NINV   & Levels & CMV & CINV & TMV & TINV &TG &Max  \\
         &           &        &       &       &        &      &     &      &  & &level\\\hline

\cite{rezaee} & $F_{1}=\sum_{}{} m_{2},m_{4},m_{6},$ & $F_{1}=$M(M(M(1,C,B'),1,A),M(A,B,1)',B)' & 4 & 3 & 3 & 4 & 4 & 8 & 6 &14 & 3\\
         & $F_{2}=\sum_{}{} m_{0},m_{1},m_{3},m_{6},$ & $F_{2}=$M(M(A,B,1)',M(A',B,C'),M(1,A,C)) & 4 & 3 & 2 & & & & & &\\
         & $F_{3}=\sum_{}{} m_{0},m_{3},m_{6}$ & $F_{2}=$M(M(M(1,C,B'),1,A),M(A',B,C'),1') & 4 & 4 & 3 & & & & & &\\\hline
         
Proposed & $F_{1}=\sum_{}{} m_{2},m_{4},m_{6},$ & $F_{1}=$M(0,C',M(1,B,A)), 
              & \textbf{2}  & \textbf{1} & \textbf{2} & 3 & 1 & \textbf{7} & 6 & \textbf{13} & \textbf{2}\\
approach       & $F_{2}=\sum_{}{} m_{0},m_{1},m_{3},m_{6}$ & $F_{2}=$M(M(C,A,B),M(1,B,A)'M(C,0,A)')& 4 & 2 & 2 & & & & & &\\
       & $F_{3}=\sum_{}{} m_{0},m_{3},m_{6}$ & $F_{3}=$M(M(C,B,A),M(1,A,B)',M(B,A',C'))& 4 & 3 & \textbf{2} & & & & & & \\\hline

\end{tabular}
\end{center}
\end{table*}

\begin{table*}[!h]
\footnotesize
%% increase Table row spacing, adjust to taste
\renewcommand{\arraystretch}{1.4}
\setlength{\tabcolsep}{.1em}
% if using array.sty, it might be a good idea to tweak the value of
% \extrarowheight as needed to properly center the text within the cells
\begin{center}
\caption{Simulation result and comparison of a 3 input/4 output function}
\label{Table4}
%% Some packages, such as MDW tools, offer better commands for making Tables
%% than the plain LaTeX2e tabular which is used here.
\begin{tabular}{|c|c|c|c|c|c|c|c|c|c|c|c|}
\hline
Approach & Minterms & Circuit & NMV  & NINV   & Levels & CMV & CINV & TMV & TINV &TG &Max  \\
         &           &        &       &       &        &      &     &      &  & &level\\\hline

\cite{rezaee} & $F_{1}=\sum_{}{} m_{1},m_{4},m_{5},m_{7},$ & $F_{1}=$M(A,B',C) & 1 & 1 & 1 & 3 & 4 & 9 & 6 &15 & 3\\
         & $F_{2}=\sum_{}{} m_{3},m_{4},m_{6},$ & $F_{2}=$M(M(M(C,B,A),1,A),A',M(A,1',C') & 4 & 3 & 3 & & & & & &\\
         & $F_{3}=\sum_{}{} m_{0},m_{2},m_{5},m_{6},$ & $F_{3}=$M(M(A,1,C),M(1',B,C),M(C,B,A)')' & 4 & 3 & 2 & & & & & &\\
         & $F_{4}=\sum_{}{} m_{4},m_{6},m_{7},$ & $F_{4}=$M(M(A,B',C),B,M(A,1',C')) & 3 & 3 & 2 & & & & & &\\\hline
         
Proposed & $F_{1}=\sum_{}{} m_{1},m_{4},m_{5},m_{7},$ & $F_{1}=$M(C,B',A)
              & 1 & 1 & 1 & 4 & 2 & 9 & 6 & 15 & 3\\
approach       & $F_{2}=\sum_{}{} m_{3},m_{4},m_{6},$ & $F_{2}=$M(M(0,C',A),M(A,C,0)',M(B,C,0))& 4 & 2 & 2 & & & & & &\\
       & $F_{3}=\sum_{}{} m_{0},m_{2},m_{5},m_{6},$& $F_{3}=$M(M(M(C,0,A)',A,B)'),B,& 5 & 4 & 3 & & & & & & \\
       & & M(M(B,0,C)',A,C')&&&&&&&&&\\
& $F_{4}=\sum_{}{} m_{4},m_{6},m_{7},$ & $F_{2}=$M(A,M(0,C',A),M(B,C,0))& 3 & 1 & 2 & & & & & &\\\hline
      
\end{tabular}
\end{center}
\end{table*}

\begin{table*}[!h]
\footnotesize
%% increase Table row spacing, adjust to taste
\renewcommand{\arraystretch}{1.4}
\setlength{\tabcolsep}{.1em}
% if using array.sty, it might be a good idea to tweak the value of
% \extrarowheight as needed to properly center the text within the cells
\begin{center}
\caption{Simulation result and comparison of a 4 input/4 output function}
\label{Table5}
%% Some packages, such as MDW tools, offer better commands for making Tables
%% than the plain LaTeX2e tabular which is used here.
\begin{tabular}{|c|c|c|c|c|c|c|c|c|c|c|c|}
\hline
Approach & Minterms & Circuit & NMV  & NINV   & Levels & CMV & CINV & TMV & TINV &TG &Max  \\
         &           &        &       &       &        &      &     &      &  & &level\\\hline

\cite{rezaee} & $F_{1}=\sum_{}{} m_{3},m_{4},m_{7},m_{15},$ & $F_{1}=$M(M(C,1',D),M(D,M(C,B,A)',B),D') & 4 & 3 & 3 & 8 & 6 & 9 & 8 & 17 & 4\\
&& &&&&&&&&&\\
         & $F_{2}=\sum_{}{} m_{1},m_{3},m_{4},m_{9},$ & $F_{2}=$M(M(1',M(D',A,1),B),M(D, & 5 & 4 & 3 & & & & & &\\
         & $m_{13},m_{15}$& M(C,B,A)',B),B') &&&&&&&&&\\
         & $F_{3}=\sum_{}{} m_{3},m_{6},m_{7},m_{11},$ & $F_{3}=$M(C,D,M(1',M(D',A,1),B)) & 3 & 2 & 3 & & & & & &\\
         & $m_{13},m_{14},m_{15}$&&&&&&&&&&\\
         & $F_{4}=\sum_{}{} m_{2},m_{6},m_{10},$ & $F_{4}=$M(M(M(C,1',D),M(D,M(C,B,A)', & 5 & 5 & 4 & & & & & &\\
 &$m_{11},m_{14}$ & B),D')',1',C)&&&&&&&&&\\\hline        
\cite{tehrani} & $F_{1}=\sum_{}{} m_{3},m_{4},m_{7},m_{15},$ & $F_{1}=$M(M(A',B,D),M(1,C,D)'),M(0,C,D)) & 4 & 2 & 2 & 3 & 2 & 12 & 5 &17 & 3\\
         & $F_{2}=\sum_{}{} m_{1},m_{3},m_{4},m_{9},$ & $F_{2}=$M(M(A',B,D),M(B,C,D)',M(0,D, & 5 & 3 & 3 & & & & & &\\
         & $m_{13},m_{15}$& M(A,B',D)))&&&&&&&&&\\
         & $F_{3}=\sum_{}{} m_{3},m_{6},m_{7},m_{11},$ & $F_{3}=$M(0,M(1,A,C),M(B,C,D)) & 3 & 0 & 2 & & & & & &\\
         & $m_{13},m_{14},m_{15}$&&&&&&&&&&\\
         & $F_{4}=\sum_{}{} m_{2},m_{6},m_{10},$ & $F_{4}=$M(C,M(A',B,D),M(0,C,D')) & 3 & 2 & 2 & & & & & &\\
         &$m_{11},m_{14}$&&&&&&&&&&\\\hline

Proposed  & $F_{1}=\sum_{}{} m_{3},m_{4},m_{7},m_{15},$ & $F_{1}=$M(M(B,A,M(B,D,C))',M(0,D,C),B) & 4 & \textbf{1} & 3 & 5 & 1 & \textbf{10} & \textbf{4} &\textbf{14} & \textbf{3}\\
    approch     & $F_{2}=\sum_{}{} m_{1},m_{3},m_{4},m_{9},$ & $F_{2}=$M(M(M(D,B,C),B,A)',B,M(M(D,0,A),B',D)) & 5 & \textbf{2} & 3 & & & & & &\\
         & $m_{13},m_{15}$&&&&&&&&&&\\
         & $F_{3}=\sum_{}{} m_{3},m_{6},m_{7},m_{11},$ & $F_{3}=$M(M(B,C,D),M(D,A,0),C)& 3 & 0 & 2 & & & & & &\\
         & $m_{13},m_{14},m_{15}$&&&&&&&&&&\\
         & $F_{4}=\sum_{}{} m_{2},m_{6},m_{10},$ & $F_{4}=$M(D',M(M(D,A,0),B',0),C) & 3 & 2 & 3 & & & & & &\\
         & $m_{11},m_{14}$&&&&&&&&&&\\\hline
      
\end{tabular}
\end{center}
\end{table*}

\section{Simulation results and analysis}
The proposed algorithm is implemented using Matlab. Simulation has been performed with a population size of $200$ chromosomes. $10\%$ of the total chromosomes in the current population pool with better fitness are selected as elite chromosomes. The process iterates till a maximum of $5000$ generations. Initially, $70\%$ of the next generation chromosomes excluding elite chromosomes are generated through crossover and leftover chromosomes are generated using mutation. Tournament selection method with size $3$ is used to select parent chromosomes for creating new offspring.  After generating the first valid chromosome corresponding to the SOP expression, the mutation rate is reduced through adopting a higher crossover rate of $80\%$. In case the valid solution is generated early before approaching the maximum generation than a dynamic approach is used to terminate the process. Once a valid chromosome is generated a periodic checking is performed, if no improvement in terms of best fitness is gained throughout 300 consecutive generations then it stops execution. 
\par Simulation has been performed with some standard functions. The simulation results and comparison using best available techniques are presented in Table [\ref{Table1}-\ref{Table5}]. Presented results demonstrate that most of the time the proposed method outperforms the existing best techniques. The important parameters used for comparison are the total majority (TMV), total inverter (TINV), total gates (TG) and maximum level.
\\
\hrule width .44\textwidth
%\begin{onethirdspacing}
%\begin{footnotesize}
\noindent NMV Number of majority\\
NIMV Number of inverter\\
CMV Common majority\\
CINV Common inverter\\
TMV Total majority\\
TINV Total inverter\\
TG Total gates
%\end{footnotesize}
%\end{onethardspacing}
\hrule width .44\textwidth
\vspace*{.08in}
In Table \ref{Table1} the combined output result using the proposed method is same as \cite{tehrani} but improvement is achieved in terms of no. of the inverter for output function $F_{1}$. In Table \ref{Table2} the proposed technique provides a much better result for the combined output as well as for the individual output functions. Our technique yields superior results for the functions presented in \ref{Table3} and \ref{Table5} as well. For the function presented in \ref{Table4} the proposed method produces similar results for the combined output as in \cite{rezaee} and considering the individual output function it is difficult to decide which one is better. However, on average the proposed technique exhibits better results than the best available techniques, which is satisfactory. Simulation is performed up to 4 outputs 4 variable functions, without loss of generality the technique is applicable to any multi-output functions with arbitrary no. of input variables.

\section{Conclusion}
In this paper, an improved technique is proposed for the automatic synthesis of optimal QCA circuits. Genetic algorithm is utilized for achieving the goal. As multi-output circuits are more common in practice, so the main focus is given to optimizing multi-output QCA circuit in particular. Optimization is performed in terms of area consumption as well as the worst-case delay of a  QCA circuit. Area consumption is optimized by minimizing no. of gates and worst case delay is optimized by minimizing the maximum level in the circuit. Importance is assigned to level and gates according to their cost in the circuit. As optimizing individual circuit may not lead to the globally optimal solution, so fitness is calculated in such way that benefited the local as well as the global optimal solution. Elitism is adopted in GA for preserving the best solutions. Improved techniques are applied in different steps of GA to get the best result. Redundancy elimination is performed for avoiding the chance of any redundancy in the final solution. Some useful strategies are introduced to reduce no. of computations. The experiment is performed using some standard functions. Simulation results signify, the proposed technique achieved better performance in terms of optimization in comparison to present best techniques. It is worth to say that the proposed technique can provide a good optimal result for any multi-output function with arbitrary no. of inputs.

\balance
\bibliography{jettaref}
\bibliographystyle{IEEEtran}

\end{document}